\documentclass[epj]{svjour}
\usepackage{graphics}
\usepackage{dcolumn}
\usepackage{bm}
\usepackage{multirow}
\usepackage{array}
\usepackage{pbox}
\usepackage{booktabs}
\usepackage[numbers]{natbib}
\usepackage{mathtools}

\usepackage{amsfonts}       
\usepackage{nicefrac}       
\usepackage{microtype}      
\usepackage{lipsum}
\usepackage{graphicx}
\usepackage{amsmath}

\usepackage{url}
\begin{document}
\title{Adversarially Learned Anomaly Detection on CMS Open Data: re-discovering the top quark}

\author{O.~Knapp\inst{1} 
  \and O.~Cerri\inst{2} \and G.~Dissertori\inst{1} \and T.~Q.~Nguyen\inst{2}  \and M.~Pierini\inst{3} \and JR.~Vlimant\inst{2}}

  
\institute{Institute for Particle Physics and Astrophysics, \\ ETH Zurich CH-8093 Zurich, Switzerland \and
  Division of Physics, Mathematics and Astronomy, \\ California Institute of Technology, 1200 E.\ California Blvd, Pasadena, CA 91125 \and
  Experimental Physics Department, European Organization for Nuclear Research (CERN), \\ CH-1211 Geneva, Switzerland \\ \email{maurizio.pierini@cern.ch}}%

\abstract{We apply an Adversarially Learned Anomaly Detection (ALAD) algorithm to the problem of detecting new physics processes in proton-proton collisions at the Large Hadron Collider. Anomaly detection based on ALAD matches performances reached by Variational Autoencoders, with a substantial improvement in some cases. Training the ALAD algorithm on $4.4$~fb$^{-1}$ of 8~TeV CMS Open Data, we show how a data-driven anomaly detection and characterization would work in real life, re-discovering the top quark by identifying the main features of the $t \bar t$ experimental signature at the LHC.}

\maketitle

\section{Introduction}

CERN's Large Hadron Collider (LHC) delivers proton-proton collisions in unique experimental conditions. Not only it accelerates protons to an unprecedented energy (6.5~TeV for each proton beam). It also operates at the highest collision frequency, producing one proton-beam crossing (an "event") every 25 nsec. By recording the information of every sensor, the two LHC multipurpose detectors, ATLAS and CMS, generate ${\cal O}(1)$~MB of data for each interesting event. The LHC big data problem consists in the incapability of storing each event, that would require writing ${\cal O}(10)$~TB/sec. In order to deal with this limitation, a typical LHC detector is equipped with a real-time event selection system, the trigger. 

The need of a trigger system for data taking has important consequences downstream. In particular, it naturally shapes the data analysis strategy for many searches for new physics phenomena (new particles, new forces, etc.) into a hypothesis test~\cite{ATLAS:2011tau}: one specifies a signal hypothesis upfront (in fact, as upfront as the trigger system) and tests the presence of the predicted kind of events (the experimental signature) on top of a background from known physics processes, against the alternative hypothesis (i.e., known physics processes alone). From a data-science perspective, this corresponds to a supervised strategy. This procedure was very successful so far, thanks to well established signal hypotheses to test (e.g., the existence of the Higgs boson). On the other hand, following this paradigm didn't produce so far any evidence for new-physics signals. While this is teaching us a lot about our Universe,~\footnote{For instance, the amount of information derived from this large number of "unsuccessful" searches has put to question the concept of "natural" new physics models, such as low scale supersymmetry. Considering that the generally prevailing pre-LHC view of particle physics was based on two pillars (the Higgs boson and low-scale natural supersymmetry), these experimental results are shaping our understanding of microscopic physics as much as, and probably even more than, the discovery of the Higgs boson.} it also raises questions on the applied methodology.

Recent works have proposed strategies, mainly based on Machine Learning (ML), to relax the underlying assumptions of a typical experimental analysis~\cite{Weisser:2016cnc,vae_lhc,DAgnolo:2018cun,DeSimone:2018efk,Farina:2018fyg,Collins:2018epr,Blance:2019ibf,Hajer:2018kqm,Heimel:2018mkt,Collins:2019jip,DAgnolo:2019vbw}, extending traditional unsupervised searches performed at colliders~\cite{Aaron:2008aa,Aaltonen:2008vt,Abazov:2011ma,CMS-PAS-EXO-14-016,Aaboud:2018ufy,Nachman:2020lpy,Andreassen_2020,Amram:2020ykb}. While many of these works focus on {\it off-line} data analysis, Ref.~\cite{vae_lhc} advocates the need to also perform an {\it on-line} event selection with anomaly detection techniques, in order to be able to save a fraction of new physics events even
when the underlying new physics scenario was unforeseen (and no dedicated trigger algorithm was put in place). Selected anomalous events could then be visually inspected (as done with the CMS exotica hotline data stream on early LHC runs in 2010-2012) or be given as input to off-line unsupervised analyses, following any of the strategies suggested in literature.

In this paper, we extend the work of Ref.~\cite{vae_lhc} in two directions: (i) we identify anomalies using an Adversarially Learned Anomaly Detection (ALAD) algorithm~\cite{alad}, which combines the strength of generative adversarial networks~\cite{gan,anogan} with that of autoencoders~\cite{kingma2014auto,an2015variational,ae_anomaly}; (ii) we demonstrate how the anomaly detection would work in real life, using the ALAD algorithm to re-discover the top quark. To this purpose we use real LHC data, released by the CMS experiment on the CERN Open Data portal~\cite{cms-opendata}. Our implementation of the ALAD model in TensorFlow~\cite{TF}, derived from the original code of Ref.~\cite{alad}, is available on GitHub~\cite{alad_lhc_code}. 

This paper is structured as follows: the ALAD algorithm is described in Section~\ref{sec:alad}. Its performance is assessed in Section~\ref{sec:alad_topclass}, repeating the study of Ref.~\cite{vae_lhc}. In Section~\ref{sec:top} we use an ALAD algorithm to re-discover the top quark on a fraction of the CMS 2012 Open Data (described in Appendix~\ref{sec:cms_opendata}). Conclusions are given in Section~\ref{sec:conclusion}.


\section{Adversarially Learned Anomaly Detection}
\label{sec:alad}

\begin{figure}[t!]
    \centering
    \includegraphics[width=\textwidth]{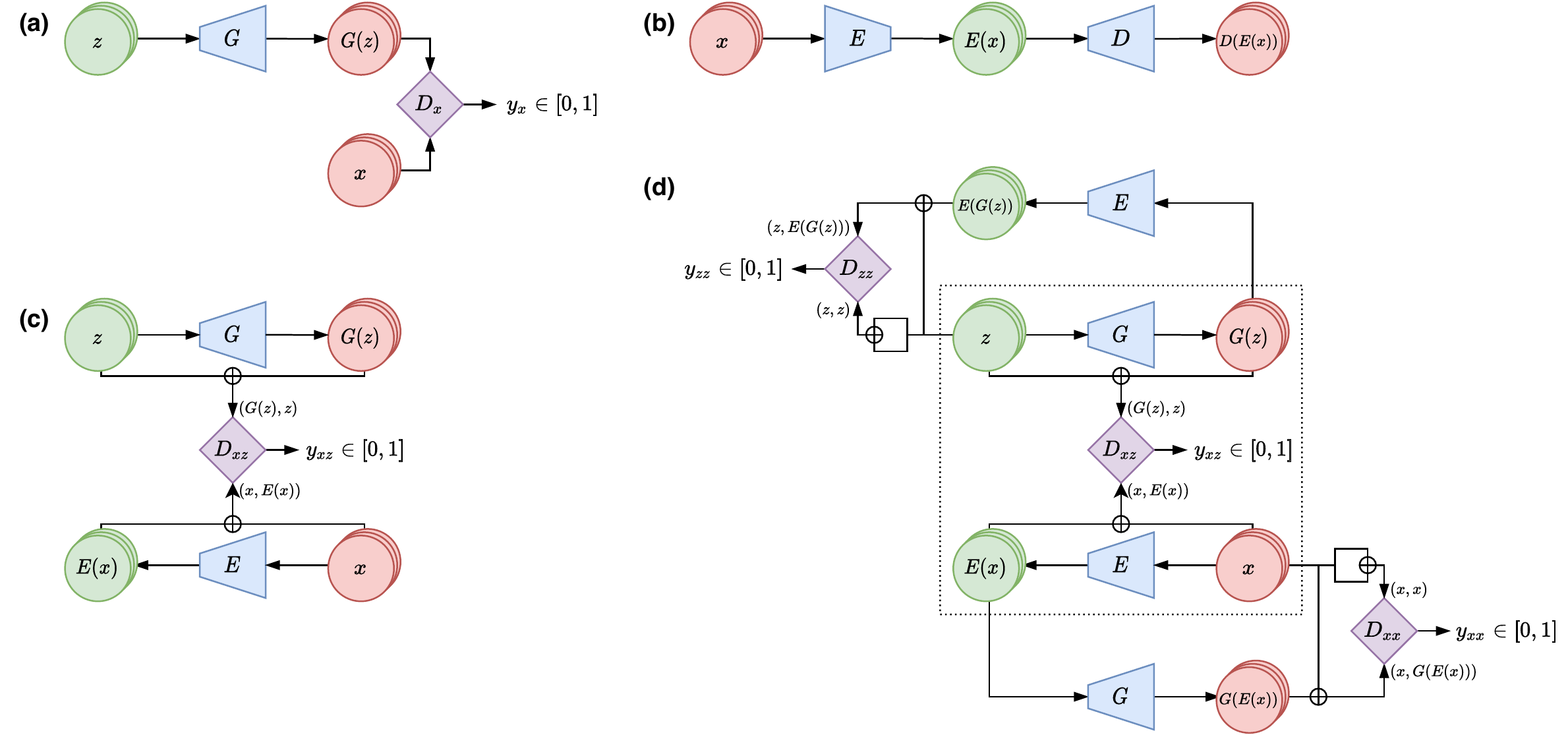}
    \caption{Comparison of Deep Network architectures: (a) In a GAN, a generator G returns samples $G(z)$ from latent-space points $z$, while a discriminator $D_x$ tries to distinguish the generated samples $G(z)$ from the real samples $x$. (b) In an autoencoder, the encoder $E$  compresses the input $x$ to a latent-space point $z$, while the decoder $D$ provides an estimate $D(z)=D(E(x))$ of $x$. (c) A  BiGAN  is built by adding to a GAN an encoder to learn the $z$ representation of the true $x$, and using the information both in the real space $\mathcal{X}$ and the latent space $\mathcal{Z}$ as input to the discriminator. (d) The ALAD model is a BiGAN in which two additional discriminators help converging to a solution which fulfils the cycle-consistency conditions $G(E(x))\approx x$ and $E(G(z))\approx z$. The $\oplus$ symbol in the figure represents a vector concatenation.
    \label{fig:architecture}}
\end{figure}

The ALAD algorithm is a kind of Generative Adversarial Network (GAN)~\cite{gan} specifically designed for anomaly detection. The basic idea underlying GANs is that two artificial neural networks compete against each other during training, as shown in Fig.~\ref{fig:architecture}. One network, the generator $G:\mathcal{Z} \rightarrow \mathcal{X}$, learns to generate new samples in the data space (e.g., proton-proton collisions in our case) aiming to resemble the samples in the training set. The other network, the discriminator $D_x: \mathcal{X} \to [0, 1]$, tries to distinguish real samples from generated ones, returning the score of a given sample to be real, as opposed of being generated by $G$. Both $G$ and $D_x$ are expressed as neural networks, which are trained against each other in a saddle-point problem:
\begin{equation}
    \underset{G}{\textup{min}}\, \underset{D_x}{\textup{max}}\, 
    \mathbb{E}_{x\sim p_\mathcal{X}}[\log\,D_x(x)] + \mathbb{E}_{z\sim p_\mathcal{Z}}[\log\,(1-D_x(G(z))]~,
    \label{eq:gan}
\end{equation}
where $p_\mathcal{X}(x)$ is the distribution over the data space $\mathcal{X}$ and $p_\mathcal{Z}(z)$ is the distribution over the latent space $\mathcal{Z}$. The solution to this problem will have the property $p_\mathcal{X}=p_G$, where $p_G$ is the distribution induced by the generator~\cite{gan}. The training typically involves alternating gradient descent on the parameters of $G$ and $D_x$ to maximize for $D_x$ (treating $G$ as fixed) and to minimize for $G$ (treating $D_x$ as fixed).

Deep learning for anomaly detection~\cite{anogan} is usually discussed in the context of (variational) autoencoders~\cite{kingma2014auto,an2015variational}. With autoencoders (cf. Fig.~\ref{fig:architecture}), one projects the input $x$ to a point $z$ of a latent-space through an encoder network $E:\mathcal{X} \rightarrow \mathcal{Z}$. An approximation $D(z)=D(E(x))$ of the input information is then reconstructed through the decoder network, $D:\mathcal{Z} \rightarrow \mathcal{X}$. The intuition is that the decoder $D$ can only reconstruct the input from the latent space representation $z$ if $x\sim p_\mathcal{X}$. Therefore, the reconstruction for an anomalous sample, which belongs to a different distribution, would typically have a higher reconstruction loss. One can then use a metric $D_{\textup{R}}$ defining the output-to-input distance (e.g., the one used in the reconstruction loss function) to derive an anomaly-score $A$:
\begin{equation}
    A(x) \sim \textup{D}_{\textup{R}}(x, D(E(x))).
\end{equation}
While this is not directly possible with GANs, since a
generated $G(z)$ doesn't correspond to a specific $x$,
several GAN-based solutions have been proposed that would be suitable for anomaly detection, as for instance in Refs.~\cite{anogan,CreswellB16b,WuBSG16,alad,gherbi:hal-02421274}.

In this work, we focus on the ALAD method~\cite{alad}, built upon the use of bidirectional-GANs (BiGAN)~\cite{bigan}. As shown in Fig.~\ref{fig:architecture}, a BiGAN model adds an encoder $E: \mathcal{X} \rightarrow \mathcal{Z}$ to the GAN construction. This encoder is trained simultaneously to the generator. The saddle point problem in Eq.(\ref{eq:gan}) is then extended as follows:
\begin{equation}
\underset{G,E}{\textup{min}}\,\underset{D_{xz}}{\textup{max}}\,
    V(D_{xz}, E, G)
    =
\underset{G,E}{\textup{min}}\, \underset{D_{xz}}{\textup{max}}\, \mathbb{E}_{x\sim p_\mathcal{X}}[\log\,D_{xz}(x, E(x))] + \mathbb{E}_{z\sim p_\mathcal{Z}}[\log\,(1-D_{xz}(G(z), z)],
\end{equation}
where $D_{xz}$ is a modified discriminator, taking inputs from both the $\mathcal{X}$ and $\mathcal{Z}$. Provided there is convergence to the global minimum, the solution has the distribution matching property $p_E(x,z)=p_G(x,z)$, where one defines $p_E(x,z) = p_E(z|x)p_\mathcal{X}(x)$ and $p_G(x,z) = p_G(x|z)p_\mathcal{Z}(z)$~\cite{bigan}. To help reaching full convergence, the ALAD model
is equipped with two additional discriminators: $D_{xx}$ and $D_{zz}$. The former discriminator together with the value function
\begin{equation}
    V(D_{xx}, E, G)=\mathbb{E}_{x\sim p_\mathcal{X}}[\log\,D_{xx}(x, x)] + \mathbb{E}_{x\sim p_\mathcal{X}}[\log\,(1-D_{xx}(x, G(E(x)))]
\end{equation}
enforces the cycle-consistency condition $G(E(x))\approx x$. The latter is added to further regularize the latent space through a similar value function:
\begin{equation}
    V(D_{zz}, E, G)=\mathbb{E}_{z\sim p_\mathcal{Z}}[\log\,D_{zz}(z, z)] + \mathbb{E}_{z\sim p_\mathcal{Z}}[\log\,(1-D_{zz}(z, E(G(z)))],
\end{equation}
enforcing the cycle condition $E(G(z)) \approx z$. The ALAD training objective consists in solving:
\begin{equation}
    \underset{G,E}{\textup{min}}\, \underset{D_{xz}, D_{xx}, D_{zz}}{\textup{max}}\, 
    V(D_{xz},E,G) + V(D_{xx}, E, G) + V(D_{zz}, E, G)~.
\end{equation}

Having multiple outputs at hand, one can associate the ALAD algorithm to several  anomaly-score definitions. Following Ref.~\cite{alad}, we consider the following four anomaly scores:
\begin{itemize}
    \item A "Logits" score, defined as: $A_L(x)=\log(D_{xx}(x, G(E(x)))$.
    \item A "Features" score, defined as: $A_F(x)=||f_{xx}(x,x) - f_{xx}(x, G(E(x)))||_1$, where $f_{xx}(\cdot,\cdot)$ are the activation values in the last hidden layer  of $D_{xx}$.
    \item The $L_1$ distance between an input $x$ and its reconstructed output $G(E(x))$: $A_{L_1}(x)= ||x - G(E(x))||_1$.
    \item The $L_2$ distance between an input $x$ and its reconstructed output $G(E(x))$: $A_{L_2}(x)= ||x - G(E(x))||_2$.
\end{itemize}

We first apply this model to the problem described in Ref~\cite{vae_lhc}, in order to obtain a direct comparison with VAE-based anomaly detection. Then, we apply this model to real LHC data (2012 CMS Open Data), showing how anomaly detection could guide physicists to discover and characterize new processes.


\section{ALAD Performance Benchmark}
\label{sec:alad_topclass}

We consider a sample of simulated LHC collisions, pre-filtered by requiring the presence of a muon with large transverse momentum ($p_T$)~\footnote{As common for collider physics, we use a Cartesian coordinate system with the $z$ axis oriented along the beam axis, the $x$ axis on the horizontal plane, and the $y$ axis oriented upward. The $x$ and $y$ axes define the transverse plane, while the $z$ axis identifies the longitudinal direction. The azimuth angle $\phi$ is computed with respect to the $x$ axis. The polar angle $\theta$ is used to compute the pseudorapidity $\eta = -\log(\tan(\theta/2))$. The transverse momentum ($p_T$) is the projection of the particle momentum on the ($x$, $y$) plane. We fix units such that $c=\hbar=1$.} and isolated from other particles. Proton-proton collision events at a center-of-mass energy $\sqrt{s}=13$~TeV are generated with the \texttt{PYTHIA8} event-generation library \cite{pythia8.2}. The generated events are further processed with the \texttt{DELPHES} library \cite{delphes} to model the detector response. Subsequently the \texttt{DELPHES} \textit{particle-flow} (PF) algorithm is applied to obtain a list of reconstructed particles for each event, the so-called PF candidates. 

Events are filtered requiring $p_T>23$~GeV and isolation~\footnote{A definition of isolation is provided in Appendix~\ref{sec:cms_opendata}.} $\textup{Iso}<\textup{0.45}$. Each collision event is represented by 21 physics-motivated high-level features (see Ref.~\cite{vae_lhc}). These input features are pre-processed before being fed to the ALAD algorithm. The discrete quantities~\footnote{$q_l$ is the charge of the lepton; $\textup{IsEle}$ is a flag set to 1 (0) if the lepton is an electron (muon); $N_\mu$ and $N_e$ are the muon and electron multiplicities, respectively.} ($q_l, \textup{IsEle}$, $N_\mu$ and $N_e$) are represented through one-hot encoding. The other features are standardized to a zero median and unit variance. The resulting vector, containing the one-hot encoded and continuous features, has a dimension of 39 and is given as input to the ALAD algorithm.

The sample, available on Zenodo~\cite{Zenodo_mpp}, consists of the following Standard Model (SM) processes:
\begin{itemize}
    \item Inclusive $W$ boson production: $W \rightarrow \ell\nu$, with $\ell = e, \mu, \tau$ being a charged lepton~\cite{cerri_olmo_2020_3675199}.
    \item Inclusive $Z$ boson production: $Z \rightarrow \ell\ell$~\cite{cerri_olmo_2020_3675203}.
    \item Multijet production from Quantum Chromodynamic (QCD) interaction~\cite{cerri_olmo_2020_3675210}.
    \item $t\bar{t}$ production~\cite{cerri_olmo_2020_3675206}.
\end{itemize}
A SM cocktail is assembled from a weighted mixture of those four processes, with weights given by the  production cross section. This cocktail's composition is given in Table \ref{table:smmix}. 

\begin{table}[!htb]
\caption{Composition of the cocktail of SM processes. The first column gives the production cross section for each process. The trigger efficiency is the fraction of events passing a loose selection on the muon $p_T$, corresponding to an online selection in the trigger system. The acceptance is the fraction of events passing our selection criteria ($p_T>23$~GeV and $\textup{Iso}<\textup{0.45}$). Each fraction is computed with respect to the previous step. The last column gives the composition of the SM-cocktail.\label{table:smmix}}
\centering
\begin{tabular}{l|lll|l}
\hline
Process & \begin{tabular}[c]{@{}l@{}}Cross\\ section {[}nb{]}\end{tabular} & \begin{tabular}[c]{@{}l@{}}Trigger\\ efficiency\end{tabular} & Acceptance & Fraction \\ \hline
$W\rightarrow \ell\nu$   & 58                & 68\%  & 55.6\%    & 59.2\% \\
QCD & $1.6 \cdot 10^5$  & 9.6\% & 0.08\%    & 33.8\%  \\
$Z\rightarrow \ell\ell$   & 20                & 77\%  & 16\%      & 6.7\%  \\
$t\bar{t}$  & 0.7               & 49\%  & 37\%      & 0.3\%   \\
\hline                               
\end{tabular}
\end{table}

We train our ALAD model on this SM cocktail and subsequently apply it to a test dataset, containing a mixture of SM events and events of physics beyond the Standard Model (BSM). In particular, we consider the following BSM datasets, also available on Zenodo: 
\begin{itemize}
    \item A leptoquark with mass 80~GeV, decaying to a $b$ quark and a $\tau$ lepton: $LQ\rightarrow b\tau$~\cite{cerri_olmo_2020_3675196}.
    \item A neutral scalar boson with mass 50~GeV, decaying to two off-shell $Z$ bosons, each forced to decay to two leptons: $A \rightarrow 4\ell$~\cite{cerri_olmo_2020_3675159}.
    \item A scalar boson with mass 60~GeV, decaying to two $\tau$ leptons: $h^0 \rightarrow \tau\tau$~\cite{cerri_olmo_2020_3675190}.
    \item A charged scalar boson with mass 60~GeV, decaying to a $\tau$ lepton and a neutrino: $h^\pm \rightarrow \tau\nu$~\cite{cerri_olmo_2020_3675178}.
\end{itemize}

As a starting point, we consider the ALAD architecture~\cite{alad} used for the \texttt{KDD99} dataset, which has similar dimensionality as our input feature vector. In this configuration, both the $D_{xx}$ and $D_{zz}$ discriminators take as input the concatenation of the two input vectors, which is processed by the network up to the single output node, activated by a sigmoid function. The $D_{xz}$ discriminator has one dense layer for each of the two inputs. The two intermediate representations are concatenated and passed to another dense layer and then to a single output node with  sigmoid activation, as for the other discriminators. The hidden nodes of the generator are activated by ReLU functions~\cite{relu}, while Leaky ReLU~\cite{leakyRelu} are used for all the other nodes. The slope parameter of the Leaky ReLU function is fixed to 0.2. The network is optimized using the Adam~\cite{adam} minimizer and minibatches of 50 events each. The training is regularized using dropout layers in the three discriminators. 

Starting from this baseline architecture, we adjust the architecture hyperparameters one by one, repeating the training while maximizing a figure of merit for anomaly detection efficiency. We perform this exercise using as anomalies the benchmark models described in Ref.~\cite{vae_lhc} and looking for a configuration that performs well on all of them. To quantify performance, we consider both the area under the receiver operating characteristic (ROC) curve and the  positive likelihood ratio $\textup{LR}_+$. We define the $\textup{LR}_+$ as the ratio between the BSM signal efficiency, i.e., the true positive rate (TPR), and the SM background efficiency, i.e., the false positive rate (FPR). The training is performed on half of the available SM events (3.4M events), leaving the other half of the SM events and the BSM samples for validation. 
From the resulting anomaly scores, we compute the ROC curve and compare it to the results of the VAE in Ref.~\cite{vae_lhc}. We further quantify the algorithm performance considering the $\textup{LR}_+$ values corresponding to an FPR of $10^{-5}$. 

The optimized architecture, adapted from Ref.~\cite{alad}, is summarized in Table~\ref{table:benchmark-hyper}. This architecture is used for all subsequent studies. We consider as hyperparameters the number of hidden layers in the five networks, the number of nodes in each hidden layer, and the dimensionality of the latent space, represented in the table by the size of the $E$ output layer.

\begin{table}[htbp]
\caption{Hyperparameters for the ALAD algorithm. Parameters in bold have been optimized for. No Dropout layer is applied wherever a dropout rate is not specified.
\label{table:benchmark-hyper}}
\centering
\begin{tabular}{lllll}
\hline
Operation  & Units & Activation & \begin{tabular}[c]{@{}l@{}}Batch\\ Norm.\end{tabular}  & 
\begin{tabular}[c]{@{}l@{}}Dropout\\ Rate\end{tabular}  \\
\hline
\multicolumn{4}{l}{$\boldsymbol{E(x)}$} \\
\hline
Number of hidden layers & \bf{2} \\
Dense   & \bf{64} & Leaky ReLU & $\times$  &  - \\
Dense   & \bf{64} & Leaky ReLU & $\times$  &  - \\
Output  & \bf{16} & Linear     & $\times$  &  - \\
\hline
\multicolumn{4}{l}{$\boldsymbol{G(z)}$} \\
\hline
Number of hidden layers & \bf{2} \\
Dense  & $\bf{64}$ & ReLU   & $\times$ &   - \\
Dense  & $\bf{64}$ & ReLU   & $\times$ &   - \\
Output & 39        & Linear & $\times$ & -  \\
\hline
\multicolumn{4}{l}{$\boldsymbol{D_{xz}(x, z)}$}\\
\hline
Number of hidden layers & \bf{2} \\
\multicolumn{4}{l}{\textit{Only on x}} \\                           
Dense    & $\bf{128}$ & Leaky ReLU & $\surd$ &  -  \\
\multicolumn{4}{l}{\textit{Only on z}} \\ 
Dense    & $\bf{128}$ & Leaky ReLU  & $\times$ &  0.5    \\
\multicolumn{4}{l}{ \textit{Concatenate outputs}} \\
Dense    & $\bf{128}$ & Leaky ReLU & $\times$ &  0.5  \\
Output    & 1          & Sigmoid    & $\times$ &  -  \\
\hline
\multicolumn{4}{l}{$\boldsymbol{D_{xx}(x, \hat{x})}$} \\
\hline
\multicolumn{4}{l}{\textit{Concatenate x and x'}} \\
Number of hidden layers & \bf{1} \\
Dense & $\bf{128}$ & Leaky ReLU & $\times$ & 0.2   \\
Output   & 1 & Sigmoid & $\times$ &  -\\
\hline
\multicolumn{4}{l}{$\boldsymbol{D_{zz}(z, \hat{z})}$}\\
\hline
\multicolumn{4}{l}{\textit{Concatenate z and z'}} \\
Number of hidden layers & \bf{1} \\
Dense   & $\bf{128}$ & Leaky ReLU  &$\times$    &    0.2\\
Output   & 1  & Sigmoid & $\times$ & - \\ 
\hline
\\
\hline
Training Parameter  & Value \\  
\hline
Optimizer                   & \multicolumn{4}{l}{Adam~($\alpha=10^{-5}$, $\beta_{1}=0.5$)}                                                   \\
Batch size                  & \multicolumn{4}{l}{\bf 50} \\
Leaky ReLU slope            & \multicolumn{4}{l}{0.2} \\
Spectral norm & \multicolumn{4}{l}{$\surd$ }  \\
Weight, bias init. & \multicolumn{4}{l}{Xavier Initializer, Constant(0)}  \\  
\hline 
\end{tabular}
\end{table}


\begin{figure}[t!]
\centering
\includegraphics[width=0.8\textwidth]{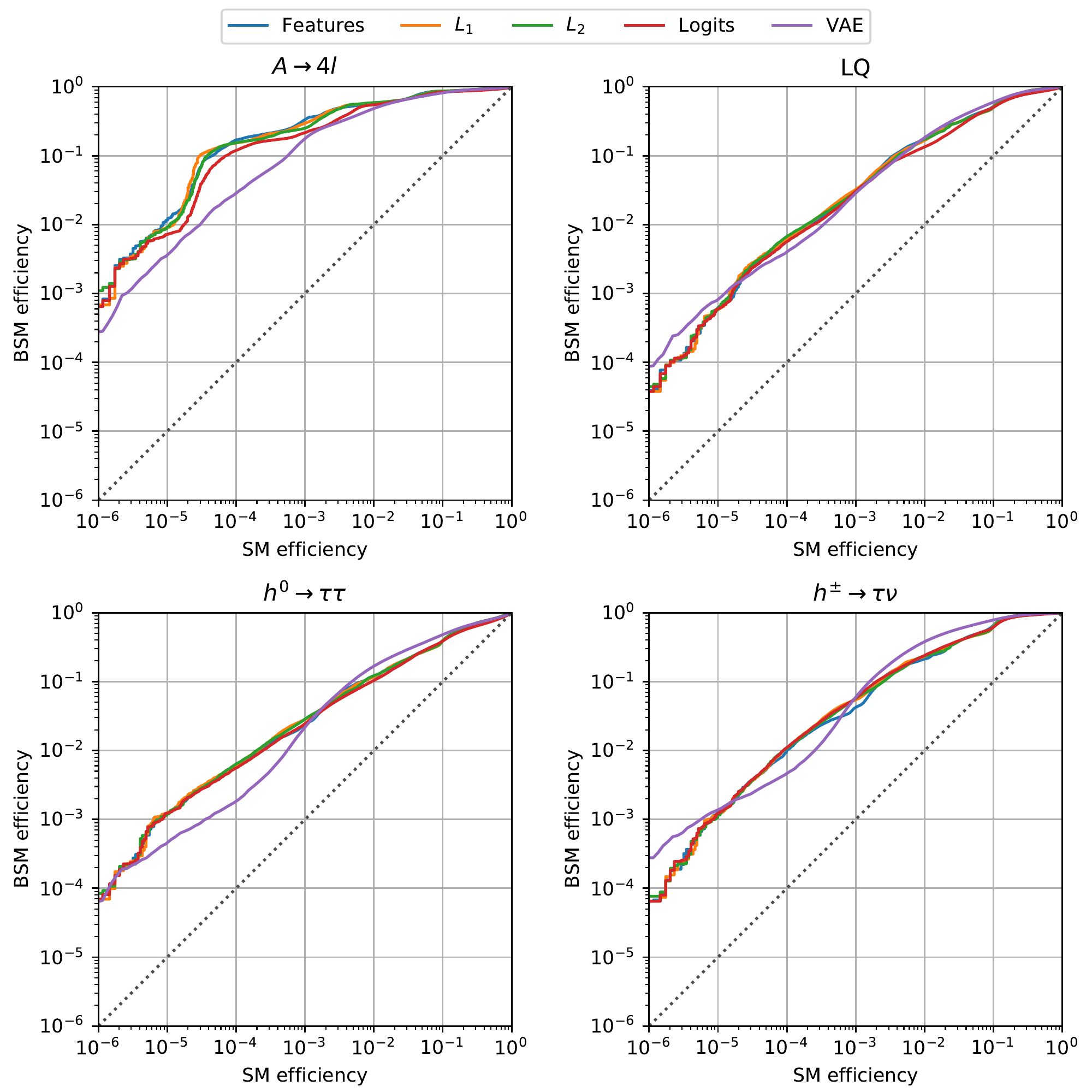}
\caption{ROC curves for the ALAD trained on the SM cocktail training set and applied to SM+BSM validation samples. The VAE curve corresponds to the best result of Ref.~\cite{vae_lhc}, which is shown here for comparison. The other four lines correspond to the different anomaly score models of the ALAD.}
\label{fig:roc}
\end{figure}

\begin{figure}[t!]
\centering
\includegraphics[width=0.6\textwidth]{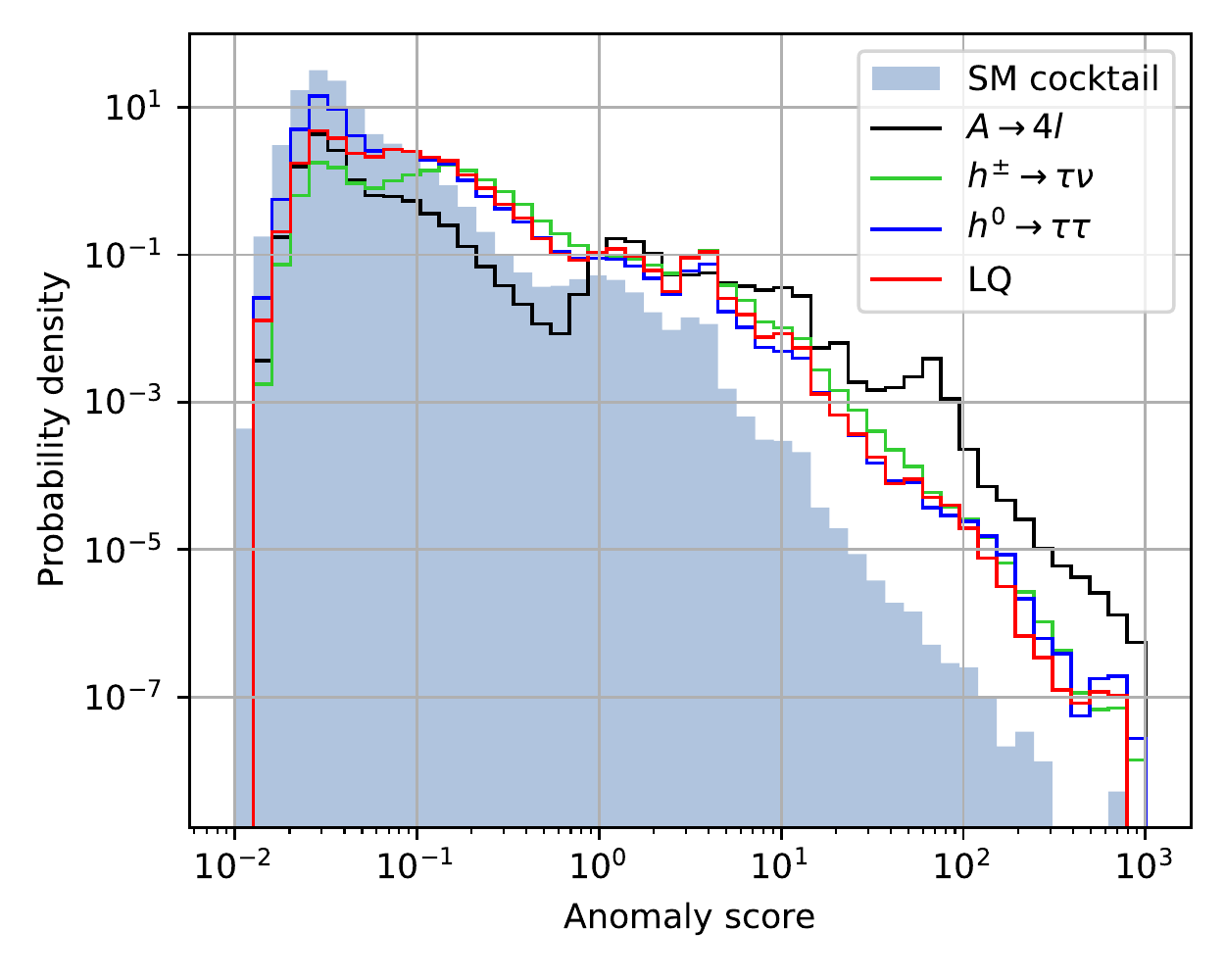}
\caption{Distribution for the $A_{L_1}$ anomaly score. The "SM cocktail" histogram corresponds to the anomaly score for the validation sample. The other four distributions refer to the scores of the four BSM benchmark models.\label{fig:score-dist}}
\end{figure}

Having trained the ALAD on the training dataset, we compute the anomaly scores for the  validation samples as well as for the four BSM samples, where each BSM process has $O(\textup{0.5M})$ samples.
Figure~\ref{fig:roc} shows the ROC curves of each BSM benchmark process, for the four considered anomaly scores. The best VAE result from Ref.~\cite{vae_lhc} is also shown for comparison. In the rest of this paper, we use the $L_1$ score as the anomaly score. Similar results would have been obtained using any of the other three anomaly scores. Figure~\ref{fig:score-dist} compares the $A_{L_1}$ distribution for each BSM process with the SM cocktail. One can clearly see that all BSM processes have an increased probability in the high-score regime compared to the SM cocktail. We further verified that the anomaly score distributions obtained on the SM-cocktail training and validation sets are consistent. This test excludes the occurrence of over-training issues. 

The ALAD algorithm  outperforms the VAE by a substantial margin on the $A \rightarrow 4\ell$ sample, providing similar performance overall, and in particular for FPR $\sim10^{-5}$, the working point chosen as a reference in Ref.~\cite{vae_lhc}. We verified that the uncertainty on the TPR at fixed FPR, computed with the Agresti–Coull interval\cite{Agresti}, is negligible when compared to the observed differences between ALAD and VAE ROC curves, i.e., the difference is statistically significant. 

\begin{figure}[t!]
\centering
\includegraphics[width=0.8\textwidth]{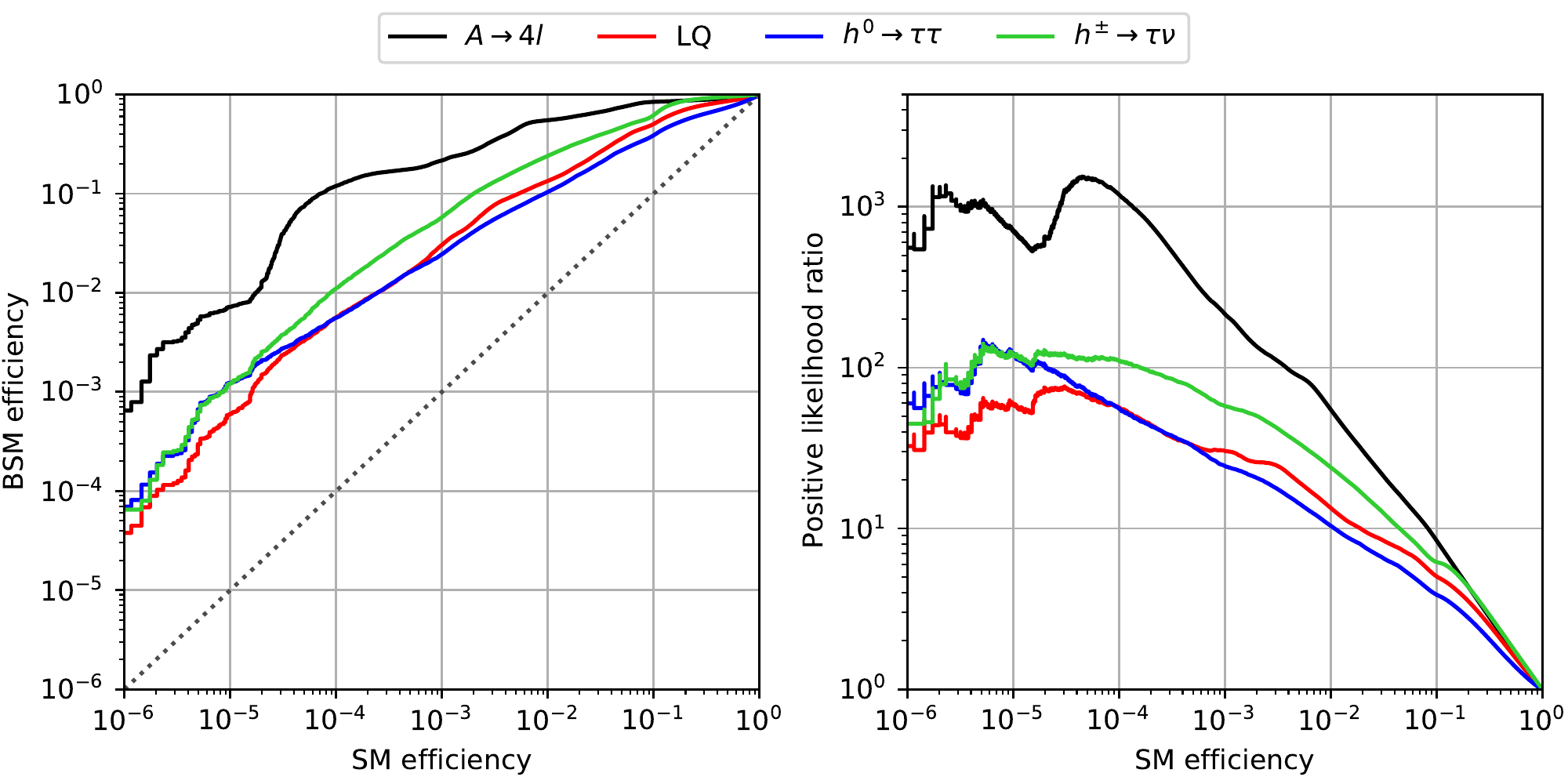}
\caption{Left: ROC curves for each BSM process obtained with the ALAD $L_1$-score model. Right: $\textup{LR}_+$ curves corresponding to the ROC curves on the left.}
\label{fig:roc-lrp}
\end{figure}

The left plot in Fig.~\ref{fig:roc-lrp} provides a comparison across different BSM models. As for the VAE, ALAD performs better on $A\rightarrow 4\ell$ and $h^{\pm} \rightarrow \tau\nu$ than for the other two BSM processes. The right plot in Figure~\ref{fig:roc-lrp} shows the $\textup{LR}_+$ values as a function of the FPR ones. The $\textup{LR}_+$  peaks at a SM efficiency of $O(10^{-5})$ for all four BSM processes and is basically constant for smaller SM-efficiency values.

\section{Re-discovering the top quark with ALAD}
\label{sec:top}

In order to test the performance of ALAD on real data, and in general to show how an anomaly detection technique could guide physicists to a discovery in a data-driven manner, we consider a scenario in which collision data from the LHC are available, but no previous knowledge about the existence of the top quark is at hand. The list of known SM processes contributing to a dataset with one isolated high-$p_T$ lepton would then include $W$ production, $Z/\gamma^*$ production and QCD multijet events, neglecting more rare processes such as diboson production. Top-quark pair production, the "unknown anomalous process", represents $\sim 1\%$ of the total dataset.

We consider a fraction of LHC events collected by the CMS experiment in 2012 and corresponding to an integrated luminosity of about 4.4~fb$^{-1}$ at a center-of-mass energy $\sqrt{s}=8$~TeV. For each collision event in the dataset, we compute a vector of physics-motivated high-level features, which we give as input to ALAD for training and inference. Details on the dataset, event content, and data processing can be found in Appendix~\ref{sec:cms_opendata}. 

We select events applying the following requirements:
\begin{itemize}
    \item At least one lepton with $p_T>23$~GeV and PF isolation $\text{Iso}<0.1$ within $|\eta|<1.4$.
    \item At least two jets with $p_T>30$ within $|\eta|<2.4$.
\end{itemize}
This selection is tuned to reduce the expected QCD multijet contamination to a negligible level, which avoids problems with the small size of the available QCD simulated dataset. This selection should not be seen as a limiting factor for the generality of the study: in real life, one would apply multiple sets of selection requirements on different triggers, in order to define multiple datasets on which different anomaly detection analyses would run. 

\begin{figure}[t!]
\centering
\includegraphics[width=0.8\textwidth]{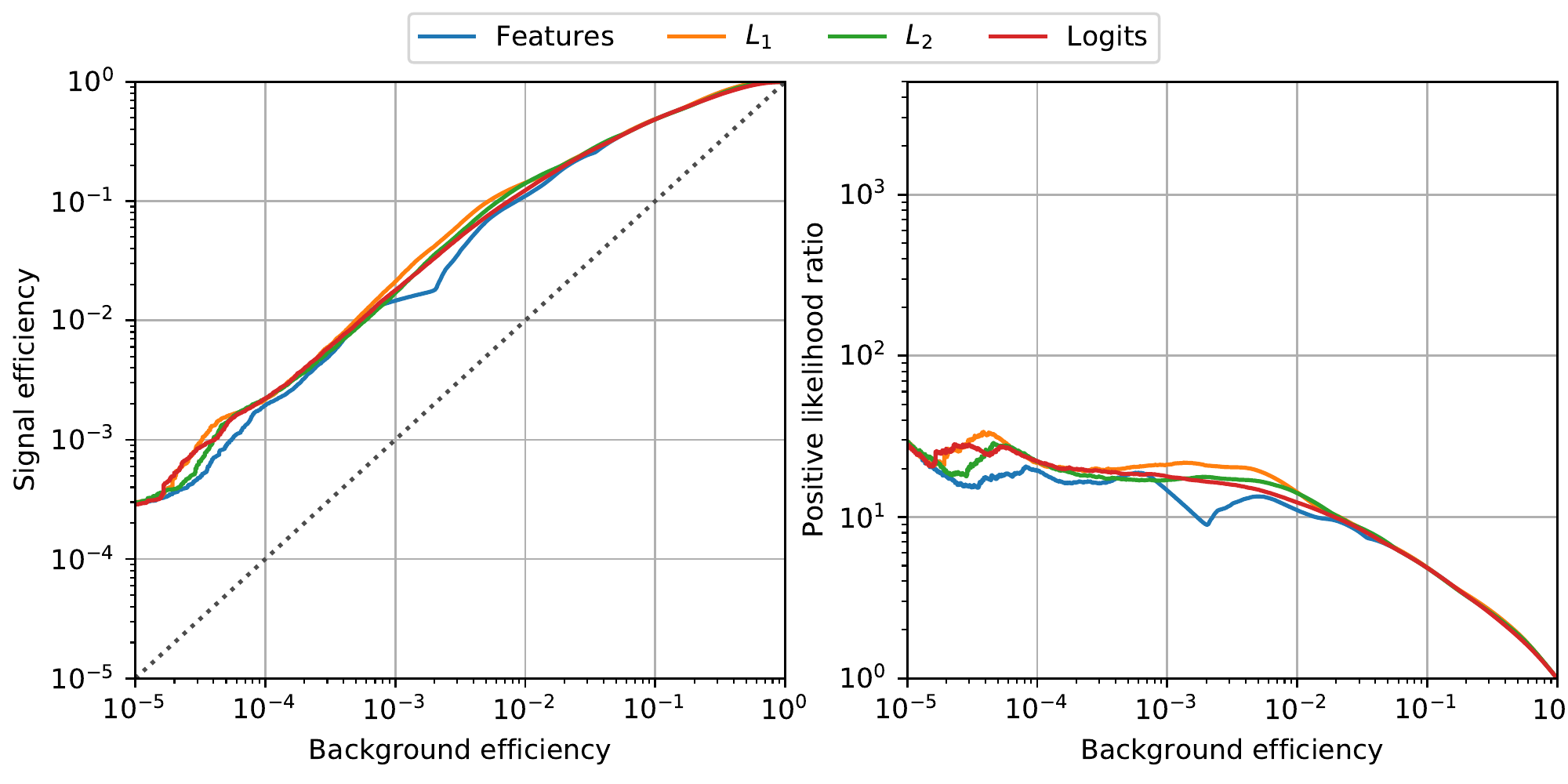}
\caption{Left: ROC curves for each anomaly score, where the signal efficiency is the fraction of $t\bar{t}$ (signal) events passing the anomaly selection, i.e., the true positive rate (TPR). The background efficiency is the fraction of background events passing the selection, i.e. the false positive rate (FPR). Right: Positive likelihood ratio ($LR_+$) curves corresponding to the ROC curves in the left.}
\label{fig:roc-tt}
\end{figure}

Our goal is to employ ALAD to isolate $t \bar t$ events as due to a rare (and pretended to be) unknown process in the selected dataset. Unlike the case discussed in Ref~\cite{vae_lhc} and Section~\ref{sec:alad_topclass}, we are not necessarily interested in a pre-defined algorithm to run on a trigger system. Given an input dataset, we want to isolate its outlier events without explicitly specifying which tail of which kinematic feature one should look at.  Because of this, we don’t split the data into a training and a validation dataset. Instead, we run the training on the full dataset. 
The training is performed fixing the ALAD architecture to the values shown in Table~\ref{table:benchmark-hyper}. In our procedure, we implicitly rely on the assumption that the anomalous events are rare, so that their modeling is not accurately learned by the ALAD algorithm.

In order to evaluate the ALAD performance, we show in Fig.~\ref{fig:roc-tt} the ROC and $\textup{LR}_+$ curves on labelled Monte Carlo (MC) simulated data, considering the $t \bar t$ sample as the signal anomaly and the SM $W$ and $Z$ production as the background. 
An event is classified as anomalous whenever the $L_1$ anomaly score is above a given threshold. The threshold is set such that the fraction of selected events is about $10^{-3}$. The anomaly selection results in a factor-20 enhancement of the $t \bar t$ contribution over the SM background, for the anomaly-defining FPR threshold. 

\begin{figure}[t!]
\centering
\includegraphics[width=0.85\textwidth]{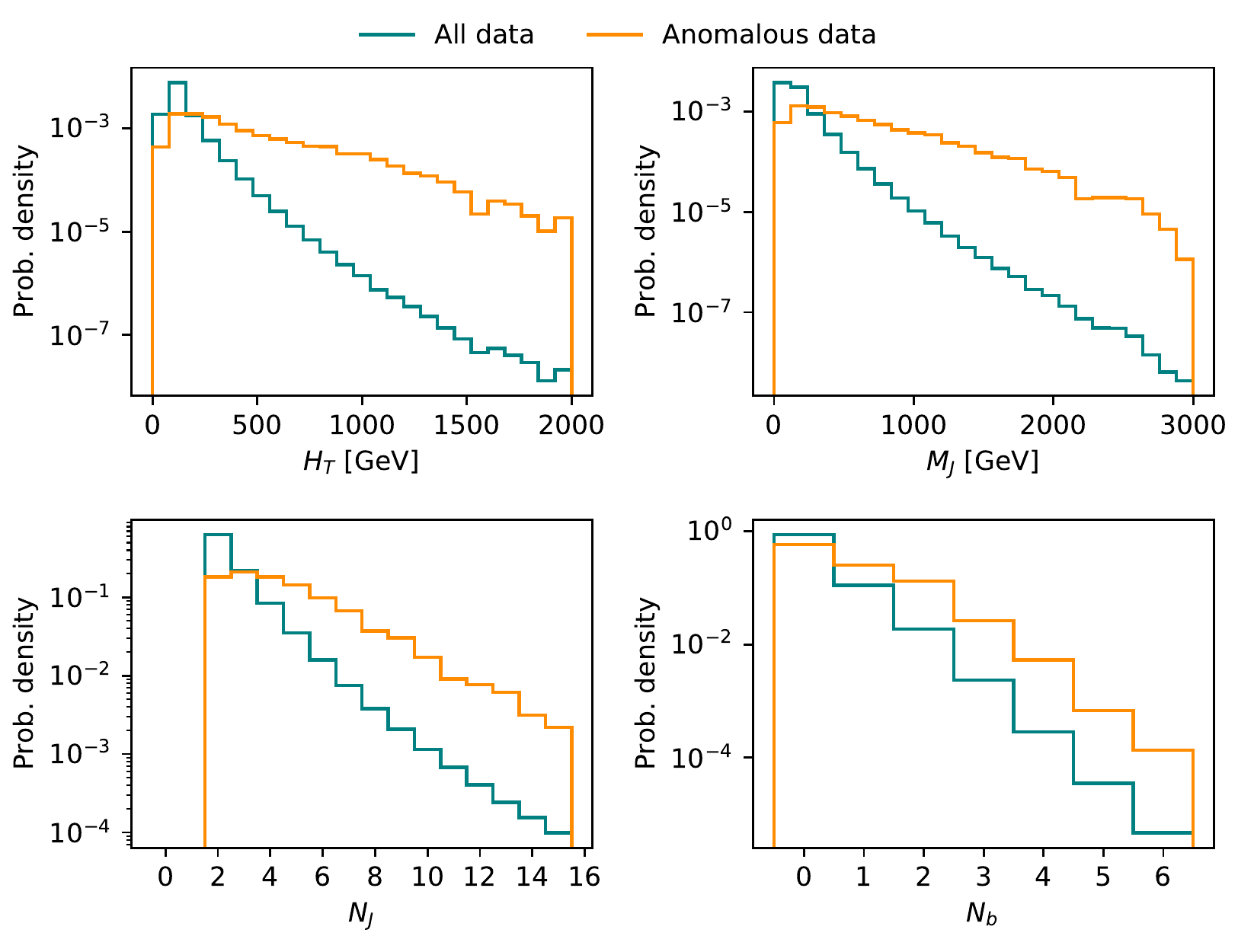}
\caption{Event distribution on data, before and after the anomaly selection: $H_T$ (top-left), $M_J$ (top-right), $N_J$ (bottom-left), and $N_b$ (bottom-right) distributions, normalized to unit area. A definition of the considered features is given in appendix~\ref{sec:cms_opendata}.\label{fig:data-dist}}
\end{figure}

\begin{figure}[p!]
\centering
\includegraphics[width=0.8\textwidth]{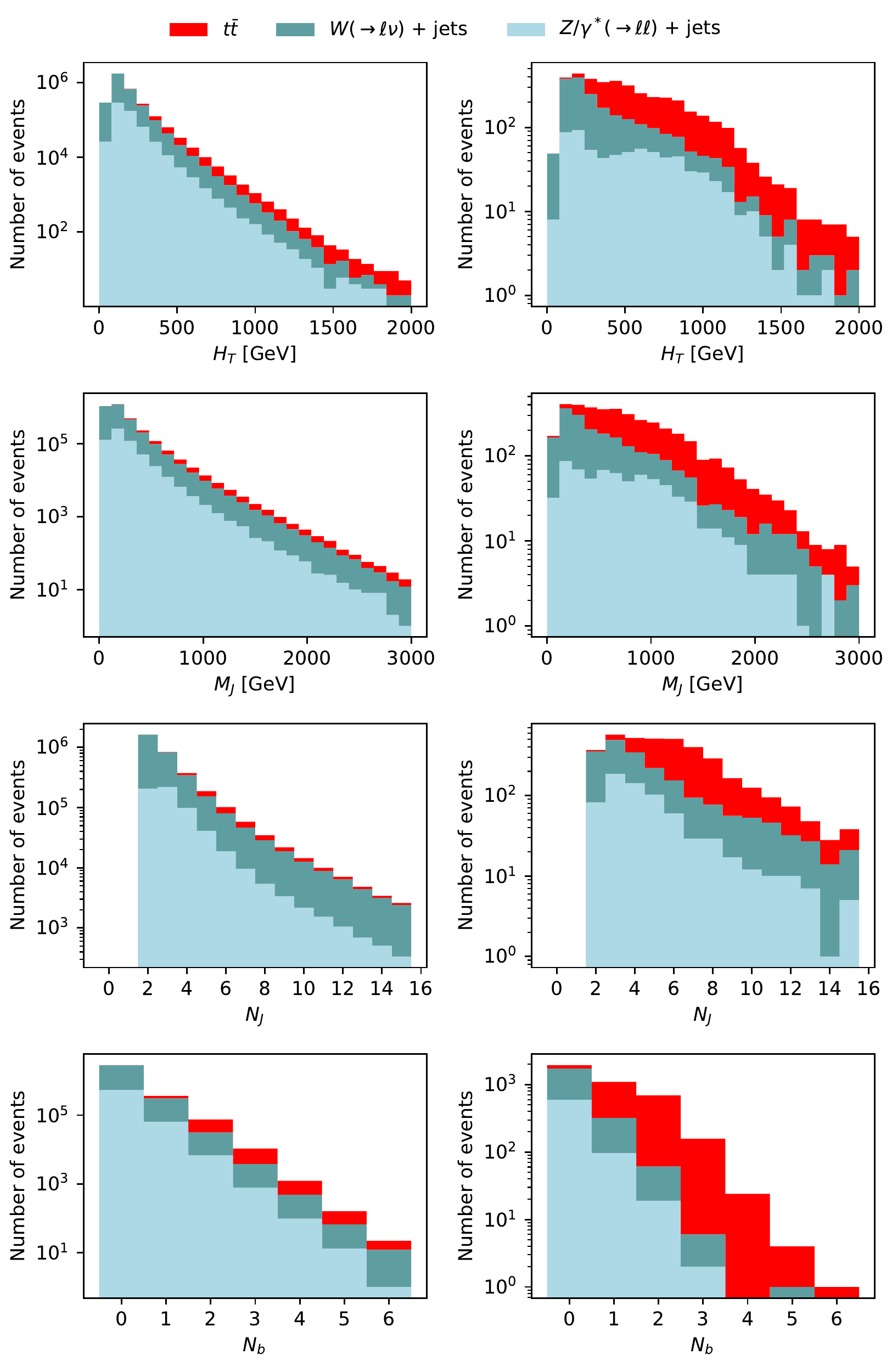}
\caption{Background event distribution from simulation before (left) and after (right) applying the anomaly selection. From top to bottom: $H_T$, $M_J$, $N_J$, and $N_b$. Distributions are normalized to an integrated luminosity $\int\!L = 4.4\,\textup{fb}^{-1}$.\label{fig:mc-dist}}
\end{figure}

Figure~\ref{fig:data-dist} shows the distributions of a subset of the input quantities before and after anomaly selection (see Appendix~\ref{sec:cms_opendata}): $H_T$, $M_J$, $N_J$, and $N_b$. These are the quantities that will become relevant in the rest of the discussion. The corresponding distributions for MC simulated events are shown in Figure~\ref{fig:mc-dist}, where the 
$t \bar t$ contribution to the selected anomalous data is highlighted. At this stage, we don't attempt a direct comparison of simulated distributions to data, since we couldn't apply the many energy scale and resolution corrections, normally applied for CMS studies. Anyhow, such a comparison is not required for our data-driven definition of anomalous events.

\begin{figure}[t!]
\centering
\includegraphics[width=0.8\textwidth]{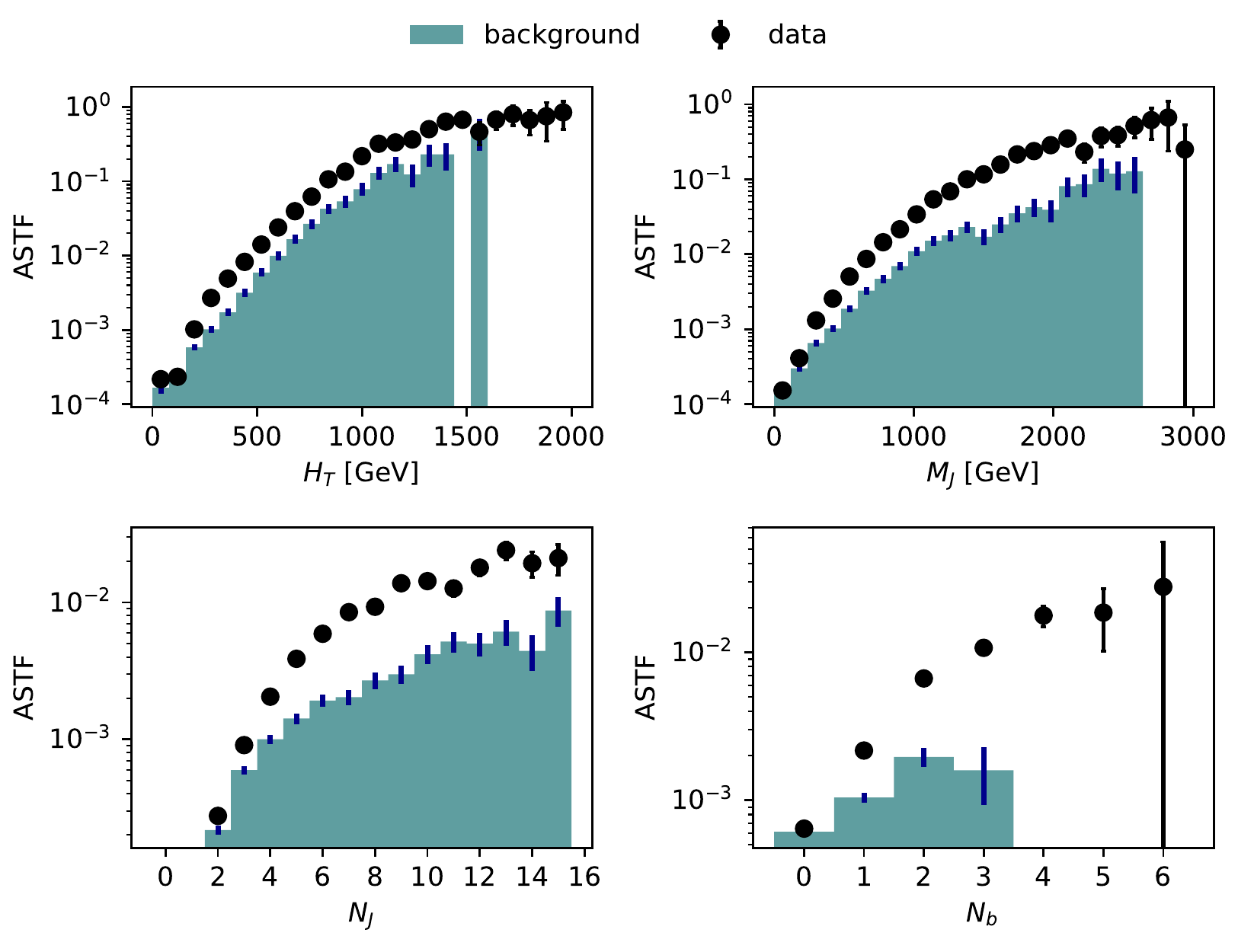}
\caption{ASTF ratios for $H_T$ (top-left), $M_J$ (top-right), $N_J$ (bottom-left), and $N_b$ (top-right) for data. 
The filled area shows the same ratio, computed on simulated background data ($W$ and $Z$ events). The uncertainty bands represent the statistical uncertainties on the ASTF ratios.\label{fig:astf}}
\end{figure}

In order to quantify the impact of the applied anomaly selection on a given quantity, we consider the bin-by-bin fraction of accepted events. We identify this differential quantity as anomaly selection transfer function (ASTF). When compared to the expectation from simulated MC data, the dependence of ASTF values on certain quantities allows one to characterize the nature of the anomaly, indicating in which range of which quantity the anomalies cluster together. 

Figure~\ref{fig:astf} shows the ASTF for the data and the simulated SM processes ($W$ and $Z$ production) defining the background data. The comparison between the data and the background ASTF suggests the presence of an unforeseen class of events, clustering at large number of jets. An excess of anomalies is observed at large jet multiplicity, which also induces an excess of anomalies at large values of $H_T$ and $M_J$. Notably, a large fraction of anomalous events has jets originating from $b$ quarks. This is the first time that a MC simulation enters our analysis. We stress the fact that this comparison between data and simulation is qualitative. At this stage, we don't need the MC-predicted ASTF values to agree with data in absence of a signal, since we are not attempting a background estimate like those performed in data-driven searches at the LHC. For us, it is sufficient to observe qualitative differences in the dependence of ASTFs on specific features. Nevertheless, a qualitative difference like those we observe in Fig.~\ref{fig:astf} could still be induced by systematic differences between data and simulation. That would still be an anomaly, but not of the kind that would lead to a discovery. In our case, we do know that the observed discrepancy is too large to be explained by a wrong modeling of $W$ and $Z$ events, given the level of accuracy reached by the CMS simulation software in Run I. In a real-life situation, one would have to run further checks to exclude this possibility.

\begin{figure}[t!]
\centering
\includegraphics[width=0.8\textwidth]{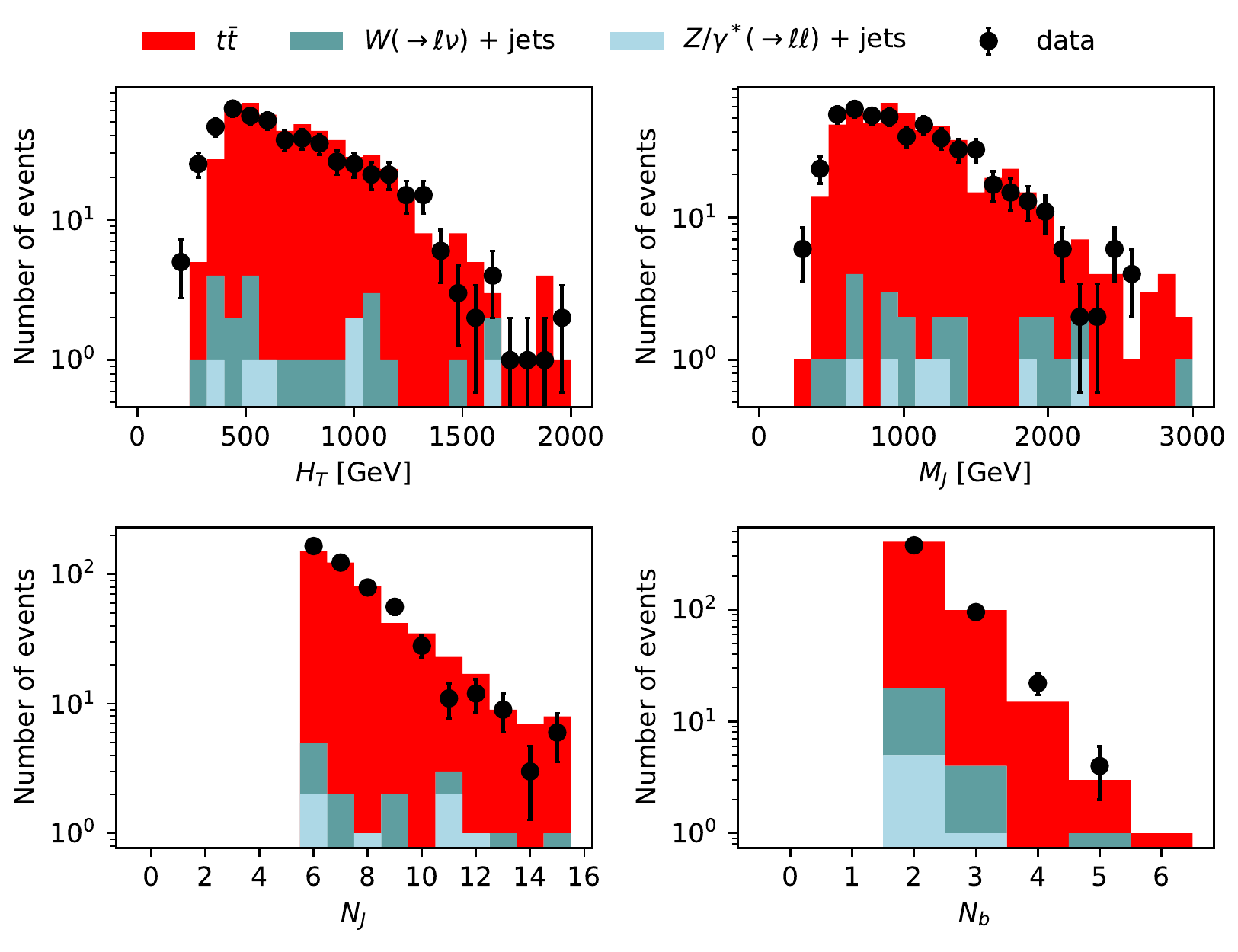}
\caption{Data distribution for $H_T$ (top-left), $M_J$ (top-right), $N_J$ (bottom-left), and $N_b$ (top-right) after the post processing selection. The filled histograms show the expectation from MC simulation, normalized to an integrated luminosity of $\int\!L = 4.4\,\textup{fb}^{-1}$), including the $t \bar t$ contribution.\label{fig:post_tt}}
\end{figure}

Starting from the ASTF plots of Fig.~\ref{fig:astf}, we define a post-processing selection (PPS), aiming to isolate a subset of anomalous events in which the residual SM background could be suppressed. In particular, we require $N_J\geq 6$ and $N_b\geq 2$. Figure~\ref{fig:post_tt} shows the distributions of some of the considered features after the PPS. According to MC expectations, almost all background events should be rejected. The same should apply to background events in data. Instead, a much larger number of events is observed. This discrepancy points to the presence of an additional process, characterized by many jets (particularly coming from $b$ jets). As a closure test, we verified that the agreement between the observed distributions in data and the expectation from MC simulation is restored once the $t \bar t$ contribution is taken into account. 

In summary, the full procedure consists of the following steps:
\begin{enumerate}
    \item Define an input dataset with an event representation which is as generic as possible.
    \item Train the ALAD (or any other anomaly detection algorithm) on it.
    \item Isolate a fraction $\alpha$ of the events, by applying a threshold on the anomaly score.
    \item Study the ASTF on as many quantities as possible, in order to define a post-processing selection that allows one to isolate the anomaly.
    \item (Optionally) once a pure sample of anomalies is isolated, a visual inspection of the events could also guide the investigation, as already suggested in Ref.\cite{vae_lhc}. 
\end{enumerate}

At this stage, one would gain an intuition about the nature of the new rare process. For instance, one could study the ASTF distribution using as a reference quantity on the $x$ axis the run period at which an event was taken. Anomalies clustering on specific run periods would most likely be due to transient detector malfunctioning or experimental problems of other nature (e.g., a bug in the reconstruction software). If instead the significant ASTF ratios point to events with a lepton, large jet multiplicity, with an excess of b-jets, one might have discovered a new heavy particle decaying to leptons and b-jets. In a world with no previous knowledge of the top quark, some bright theorist would explain the anomaly proposing the existence of a third up-type quark with unprecedented heavy mass. 

\section{Conclusions}
\label{sec:conclusion}

We presented an application of ALAD to the search for new physics at the LHC. Following the study presented in Ref.~\cite{vae_lhc}, we show how this algorithm matches (and in some cases improves) the performance obtained with variational autoencoders. The ALAD architecture also offers practical advantages with respect to the VAE of Ref.~\cite{vae_lhc}. Besides providing an improved performance in some cases, it offers an easier training procedure. Good performance can be achieved using a standard MSE loss, unlike the VAE model in Ref.~\cite{vae_lhc}, for which good performance was obtained only after a heavy customization of the loss function.

We train the ALAD algorithm on a sample of real data, obtained by processing part of the 2012 CMS Open data. On these events, we show how one could detect the presence of $t \bar t$ events, a $\sim 4\%$ population (after selection requirements are applied) pretended to originate from a previously unknown process. Using the anomaly score of the trained ALAD algorithm, we define a post-selection procedure that let us isolate an almost pure subset of anomalous events. Furthermore, we present a strategy based on ASTF distributions to characterize the nature of an observed excess and we show its effectiveness on the specific example at hand. Further studies should be carried on to demonstrate the robustness of this strategy for more rare processes. 

For the first time, our study shows with real LHC data that anomaly detection techniques can highlight the presence of rare phenomena in a data-driven manner. This result could help promoting a new class of data-driven studies in the next run of the LHC, possibly offering additional guidance in the search for new physics. 

As already stressed in Ref~\cite{vae_lhc}, the promotion of these algorithms to the trigger system of the LHC experiments could allow to mitigate for the limited bandwidth of the experiments. Unlike Ref~\cite{vae_lhc}, the strategy presented in this paper would also require a substantial fraction of normal events to be saved (through pre-scaled triggers), in order to compute the ASTF ratios with sufficient precision. In this respect, there would be a cost in terms of trigger throughput, that could be compensated by foreseeing dedicated scouting streams~\cite{scouting1,scouting2,scouting3} with reduced event content, tailored to the computation of the ASTF ratios in offline analyses. Putting in place this strategy for the LHC Run III could be an interesting way to extend the physics reach of the LHC experiments. 

\section*{Acknowledgments}

This work was possible thanks to the commitment of the CMS collaboration to release its data and MC samples through the CERN Open Data portal. We would like to thank our CMS colleagues and the CERN Open Data team for their effort to promote open access to science. In particular, we thank Kati Lassila-Perini for her precious help.

This project is partially supported by the European Research Council (ERC) under the European Union's Horizon 2020 research and innovation program (grant agreement n$^o$ 772369) and by the United States Department of Energy, Office of High Energy Physics Research under Caltech Contract No. DE-SC0011925.

This work was conducted at ``\textit{iBanks},'' the AI GPU cluster at Caltech. 
We acknowledge NVIDIA, SuperMicro and the Kavli Foundation for their support of ``\textit{iBanks}.''  

\section*{Appendix}

\appendix

\section{CMS Single Muon Open Data}
\label{sec:cms_opendata}

The re-discovery of the top quark, described in Section~\ref{sec:top}, is performed on a sample of real LHC collisions, collected by the CMS experiment and released on the CERN Open Data portal~\cite{cms-opendata}. The collision data correspond to the SingleMu dataset in the Run2012B period~\cite{data_singlemu}. This dataset results from the logic {\tt OR} of many triggers requiring a reconstructed muon. Most of the events were collected by an inclusive isolated-muon trigger, with requirements very similar to those applied as a pre-selection in Section~\ref{sec:alad_topclass}~\cite{vae_lhc}. 

Once collected, the raw data recorded by the CMS detector were processed by a global event reconstruction algorithm, based on Particle Flow (PF)~\cite{particle_flow}. This processing step gives as output a list of so-called PF candidates (electrons, muons, photons, charged hadrons and neutral hadrons), reconstructed combining measurements from different detectors to achieve maximal accuracy. 

While the ALAD training is performed directly on data, samples from MC simulation are also used in the study, in order to validate the model and its findings on a labelled sample. To this purpose, we considered samples of $W$+jets~\cite{data-9863,data-9864,data-9865}, $Z/\gamma^*$+jets~\cite{data-7719,data-7721,data-7722,data-7723}, and $t \bar t$~\cite{data-9588} events. No sample of QCD multijet events was considered, since its contribution was found to be negligible after the baseline requirements on the muon and on two additional jets (see Section~\ref{sec:top}).  These MC samples were generated by the CMS collaboration with different libraries and processed by a full Geant4 simulation~\cite{Agostinelli:2002hh}. The same reconstruction software used on data was applied to the output of the simulation, so that the same lists of PF candidates are available in this case. 


Following a procedure similar to that of Ref.~\cite{vae_lhc}, we take as input the lists of PF candidates and compute a set of physics motivated features, which are used as input to train the ALAD algorithm:

We consider 14 event-related quantities:
\begin{itemize}
    \item $H_T$ -- The scalar sum of the transverse momenta ($p_T$) of all jets having $p_T>30$~GeV and $\left | \eta \right |<2.4$. 
    \item $M_J$ -- The invariant mass of all jets entering the $H_T$ sum.
    \item $N_J$ -- The number of jets entering the $H_T$ sum.
    \item $N_B$ -- The number of jets identified as originating from a $b$ quark.
    \item $p_T^\textup{miss}$ -- The missing transverse momentum defined as $p_T^\textup{miss} = \left | - \sum_q \mathbf{p}_T^q \right |$, where the sum goes over all PF candidates.
    \item $p_{T,TOT}^\mu$ -- The vector sum of the $\mathbf{p}_T$ of all PF muons in the event having $p_T>0.5$~GeV.
    \item $M_\mu$ -- The combined invariant mass of all muons entering the above sum in $p_{T,TOT}^\mu$.
    \item $N_\mu$ -- The number of muons entering the sum in $p_{T,TOT}^\mu$.
    \item $p_{T,TOT}^e$ -- The vector sum of the $\mathbf{p}_T$ of all PF electrons in the event having $p_T>0.5$~GeV.
    \item $M_e$ -- The combined invariant mass of all electrons entering the above sum in $p_{T,TOT}^e$.
    \item $N_e$ -- The number of electrons entering the sum in $p_{T,TOT}^e$.
    \item $N_{neu}$ -- The number of all neutral hadron PF-candidates.
    \item $N_{ch}$ -- The number of all charged hadron PF-candidates.
    \item $N_{\gamma}$ -- The number of all photon PF-candidates.
\end{itemize}

We further consider 10 quantities, specific to the highest-$p_T$ lepton in the event:
\begin{itemize}
    \item $p_T^\ell$ -- The lepton $p_T$.
    \item $\eta_\ell$ -- The lepton pseudorapidity.
    \item $q_\ell$ -- The lepton charge (either $-1$ or $+1$)
    \item $\textup{Iso}_{ch}^\ell$ -- The lepton isolation, defined as the ratio between the scalar sum of the $p_T$ of the other charged PF candidates with angular distance $\Delta R= \sqrt{\Delta \eta^2 + \Delta \phi^2} < 0.3$ from the lepton, and the lepton $p_T$.
    \item $\textup{Iso}_{neu}^\ell$ -- Same as $\textup{Iso}_{ch}^\ell$ but with the sum going over all neutral hadrons.
    \item $\textup{Iso}_{\gamma}^\ell$  -- Same as $\textup{Iso}_{ch}^\ell$ but with the sum going over all photons.
    \item $\textup{Iso}$ -- The total isolation given by $\textup{Iso}=\textup{Iso}_{ch}^\ell+\textup{Iso}_{neu}^\ell+\textup{Iso}_{\gamma}^\ell$.
    \item $M_T$ -- The combined transverse mass of the lepton and the $E_T^\textup{miss}$ system, which is given by $$M_T= \sqrt{2 p_T^\ell E_T^\textup{miss} (1-\cos \Delta \phi)}.$$. 
    \item $p_{T,\parallel}^\textup{miss}$ -- The parallel component of $p_T^\textup{miss}$ with respect to the lepton.
    \item $p_{T,\perp}^\textup{miss}$ -- The orthogonal component of $p_T^\textup{miss}$ with respect to the lepton.
    \item $\textup{IsEle}$ -- A flag set to 1 if the lepton is an electron, 0 if it is a muon.
\end{itemize}

\bibliographystyle{epj}
\bibliography{biblio_epj}

\begin{thebibliography}{62}

\bibitem{ATLAS:2011tau}
{ATLAS and CMS Collaborations}, \emph{{Procedure for the LHC Higgs boson search
  combination in Summer 2011}} (2011), aTL-PHYS-PUB-2011-011, CMS-NOTE-2011-005
  \url{https://inspirehep.net/literature/1196797}

\bibitem{Weisser:2016cnc}
C.~Weisser, M.~Williams (2016), \texttt{1612.07186}

\bibitem{vae_lhc}
O.~Cerri, T.Q. Nguyen, M.~Pierini, M.~Spiropulu, J.R. Vlimant, JHEP
  \textbf{05}, 036 (2019), \texttt{1811.10276}

\bibitem{DAgnolo:2018cun}
R.T. D'Agnolo, A.~Wulzer, Phys. Rev. \textbf{D99}, 015014 (2019),
  \texttt{1806.02350}

\bibitem{DeSimone:2018efk}
A.~De~Simone, T.~Jacques, Eur. Phys. J. C \textbf{79}, 289 (2019),
  \texttt{1807.06038}

\bibitem{Farina:2018fyg}
M.~Farina, Y.~Nakai, D.~Shih, Phys. Rev. D \textbf{101}, 075021 (2020),
  \texttt{1808.08992}

\bibitem{Collins:2018epr}
J.H. Collins, K.~Howe, B.~Nachman, Phys. Rev. Lett. \textbf{121}, 241803
  (2018), \texttt{1805.02664}

\bibitem{Blance:2019ibf}
A.~Blance, M.~Spannowsky, P.~Waite, JHEP \textbf{10}, 047 (2019),
  \texttt{1905.10384}

\bibitem{Hajer:2018kqm}
J.~Hajer, Y.Y. Li, T.~Liu, H.~Wang, Phys. Rev. D \textbf{101}, 076015 (2020),
  \texttt{1807.10261}

\bibitem{Heimel:2018mkt}
T.~Heimel, G.~Kasieczka, T.~Plehn, J.M. Thompson, SciPost Phys. \textbf{6}, 030
  (2019), \texttt{1808.08979}

\bibitem{Collins:2019jip}
J.H. Collins, K.~Howe, B.~Nachman, Phys. Rev. \textbf{D99}, 014038 (2019),
  \texttt{1902.02634}

\bibitem{DAgnolo:2019vbw}
R.T. D'Agnolo, G.~Grosso, M.~Pierini, A.~Wulzer, M.~Zanetti (2019),
  \texttt{1912.12155}

\bibitem{Aaron:2008aa}
F.D. Aaron et~al. (H1), Phys. Lett. \textbf{B674}, 257 (2009),
  \texttt{0901.0507}

\bibitem{Aaltonen:2008vt}
T.~Aaltonen et~al. (CDF), Phys. Rev. \textbf{D79}, 011101 (2009),
  \texttt{0809.3781}

\bibitem{Abazov:2011ma}
V.M. Abazov et~al. (D0), Phys. Rev. \textbf{D85}, 092015 (2012),
  \texttt{1108.5362}

\bibitem{CMS-PAS-EXO-14-016}
\emph{{MUSiC, a Model Unspecific Search for New Physics, in pp Collisions at
  $\sqrt{s}=8\,\mathrm{TeV}$}} (2017), cMS-PAS-EXO-14-016
  \url{https://cds.cern.ch/record/225665}

\bibitem{Aaboud:2018ufy}
M.~Aaboud et~al. (ATLAS), Eur. Phys. J. \textbf{C79}, 120 (2019),
  \texttt{1807.07447}

\bibitem{Nachman:2020lpy}
B.~Nachman, D.~Shih, Phys. Rev. D \textbf{101}, 075042 (2020),
  \texttt{2001.04990}

\bibitem{Andreassen_2020}
A.~Andreassen, B.~Nachman, D.~Shih, Physical Review D \textbf{101} (2020)

\bibitem{Amram:2020ykb}
O.~Amram, C.M. Suarez (2020), \texttt{2002.12376}

\bibitem{alad}
H.~Zenati, M.~Romain, C.S. Foo, B.~Lecouat, V.R. Chandrasekhar (2018),
  \texttt{arXiv:1812.02288}

\bibitem{gan}
I.J. Goodfellow, J.~Pouget-Abadie, M.~Mirza, B.~Xu, D.~Warde-Farley, S.~Ozair,
  A.~Courville, Y.~Bengio (2014), \texttt{arXiv:1406.2661}

\bibitem{anogan}
T.~Schlegl, P.~Seeböck, S.M. Waldstein, U.~Schmidt-Erfurth, G.~Langs (2017),
  \texttt{arXiv:1703.05921}

\bibitem{kingma2014auto}
D.P. {Kingma}, M.~{Welling}, ArXiv e-prints  (2013), \texttt{1312.6114}

\bibitem{an2015variational}
J.~An, S.~Cho, Special Lecture on IE \textbf{2}, 1 (2015)

\bibitem{ae_anomaly}
C.~Zhou, R.C. Paffenroth, \emph{Anomaly Detection with Robust Deep
  Autoencoders}, in \emph{Proceedings of the 23rd ACM SIGKDD International
  Conference on Knowledge Discovery and Data Mining} (Association for Computing
  Machinery, New York, NY, USA, 2017), p. 665–674,
  \texttt{https://doi.org/10.1145/3097983.3098052}

\bibitem{cms-opendata}
Open Data Repository:
  \url{http://opendata.cern.ch/search?page=1&size=20&experiment=CMS}

\bibitem{TF}
M.~Abadi et~al., \emph{{TensorFlow}: Large-scale machine learning on
  heterogeneous systems} (2015), \url{http://tensorflow.org/}

\bibitem{alad_lhc_code}
O.~Knapp, \emph{Adversarially learned anomaly detection for the lhc}, gitHub:
  \url{https://github.com/oliverkn/alad-for-lhc}

\bibitem{CreswellB16b}
A.~Creswell, A.A. Bharath, CoRR \textbf{abs/1611.05644} (2016),
  \texttt{1611.05644}

\bibitem{WuBSG16}
Y.~Wu, Y.~Burda, R.~Salakhutdinov, R.B. Grosse, CoRR \textbf{abs/1611.04273}
  (2016), \texttt{1611.04273}

\bibitem{gherbi:hal-02421274}
E.~Gherbi, B.~Hanczar, J.C. Janodet, W.~Klaudel, \emph{{An Encoding Adversarial
  Network for Anomaly Detection}}, in \emph{{11th Asian Conference on Machine
  Learning (ACML 2019)}} (Nagoya, Japan, 2019), Vol. 101 of \emph{JMLR:
  Workshop and Conference Proceedings}, pp. 1--16,
  \texttt{https://hal.archives-ouvertes.fr/hal-02421274}

\bibitem{bigan}
J.~Donahue, P.~Krähenbühl, T.~Darrell (2016), \texttt{arXiv:1605.09782}

\bibitem{pythia8.2}
T.~Sjöstrand, S.~Ask, J.R. Christiansen, R.~Corke, N.~Desai, P.~Ilten,
  S.~Mrenna, S.~Prestel, C.O. Rasmussen, P.Z. Skands, Comput. Phys. Commun.
  \textbf{191} (2015)

\bibitem{delphes}
J.~de~Favereau, C.~Delaere, P.~Demin, A.~Giammanco, V.~Lemaître, A.~Mertens,
  M.~Selvaggi (DELPHES 3), JHEP \textbf{02}, 057 (2014), \texttt{1307.6346}

\bibitem{Zenodo_mpp}
M.~Pierini et~al., mPP Zenodo community:
  \url{https://zenodo.org/communities/mpp-hep/?page=1&size=20}

\bibitem{cerri_olmo_2020_3675199}
O.~Cerri, T.~Nguyen, M.~Pierini, J.R. Vlimant, \emph{New physics mining at the
  large hadron collider: $w \to \ell \nu$:
  \url{https://doi.org/10.5281/zenodo.3675199}} (2020)

\bibitem{cerri_olmo_2020_3675203}
O.~Cerri, T.~Nguyen, M.~Pierini, J.R. Vlimant, \emph{New physics mining at the
  large hadron collider: $z \to \ell \ell$:
  \url{https://doi.org/10.5281/zenodo.3675203}} (2020)

\bibitem{cerri_olmo_2020_3675210}
O.~Cerri, T.~Nguyen, M.~Pierini, J.R. Vlimant, \emph{New physics mining at the
  large hadron collider: Qcd multijet production:
  \url{https://doi.org/10.5281/zenodo.3675210}} (2020)

\bibitem{cerri_olmo_2020_3675206}
O.~Cerri, T.~Nguyen, M.~Pierini, J.R. Vlimant, \emph{New physics mining at the
  large hadron collider: $t \bar t$ production:
  \url{https://doi.org/10.5281/zenodo.3675206}} (2020)

\bibitem{cerri_olmo_2020_3675196}
O.~Cerri, T.~Nguyen, M.~Pierini, J.R. Vlimant, \emph{New physics mining at the
  large hadron collider: Lq$\to b \tau$:
  \url{https://doi.org/10.5281/zenodo.3675196}} (2020)

\bibitem{cerri_olmo_2020_3675159}
O.~Cerri, T.~Nguyen, M.~Pierini, J.R. Vlimant, \emph{New physics mining at the
  large hadron collider: $a \to 4 \ell$:
  \url{https://doi.org/10.5281/zenodo.3675159}} (2020),
  \texttt{https://doi.org/10.5281/zenodo.3675159}

\bibitem{cerri_olmo_2020_3675190}
O.~Cerri, T.~Nguyen, M.~Pierini, J.R. Vlimant, \emph{New physics mining at the
  large hadron collider: $h^0 \to \tau \tau$:
  \url{https://doi.org/10.5281/zenodo.3675190}} (2020)

\bibitem{cerri_olmo_2020_3675178}
O.~Cerri, T.~Nguyen, M.~Pierini, J.R. Vlimant, \emph{New physics mining at the
  large hadron collider: $h^+ \to \tau \nu$:
  \url{https://doi.org/10.5281/zenodo.3675178}} (2020)

\bibitem{relu}
V.~Nair, G.E. Hinton, \emph{Rectified Linear Units Improve Restricted Boltzmann
  Machines.}, in \emph{ICML} (2010),
  \texttt{https://www.cs.toronto.edu/~fritz/absps/reluICML.pdf}

\bibitem{leakyRelu}
A.L. Maas, A.Y. Hannun, A.Y. Ng, \emph{Rectifier nonlinearities improve neural
  network acoustic models}, in \emph{in ICML Workshop on Deep Learning for
  Audio, Speech and Language Processing} (2013)

\bibitem{adam}
D.P. Kingma, J.~Ba, \emph{Adam: {A} Method for Stochastic Optimization}, in
  \emph{3rd International Conference on Learning Representations, {ICLR} 2015,
  San Diego, CA, USA, May 7-9, 2015, Conference Track Proceedings} (2015),
  \texttt{http://arxiv.org/abs/1412.6980}

\bibitem{Agresti}
A.~Agresti, B.A. Coull, The American Statistician \textbf{52} (1998)

\bibitem{scouting1}
D.~Anderson, PoS \textbf{ICHEP2016}, 190 (2016)

\bibitem{scouting2}
S.~Mukherjee, \emph{{Data Scouting : A New Trigger Paradigm}}, in \emph{{5th
  Large Hadron Collider Physics Conference}} (2017), \texttt{1708.06925}

\bibitem{scouting3}
J.~Duarte, \emph{{Fast Reconstruction and Data Scouting}}, in \emph{{4th
  International Workshop Connecting The Dots 2018 (CTD2018) Seattle,
  Washington, USA, March 20-22, 2018}} (2018), \texttt{1808.00902}

\bibitem{data_singlemu}
{CMS Collaboration} (2017), \url{http://opendata.cern.ch/record/6021}

\bibitem{particle_flow}
A.~Sirunyan et~al. (CMS), JINST \textbf{12}, P10003 (2017), \texttt{1706.04965}

\bibitem{data-9863}
{CMS Collaboration} (2017), \url{http://opendata.cern.ch/record/9863}

\bibitem{data-9864}
{CMS Collaboration} (2017), \url{http://opendata.cern.ch/record/9864}

\bibitem{data-9865}
{CMS Collaboration} (2017), \url{http://opendata.cern.ch/record/9865}

\bibitem{data-7719}
{CMS Collaboration} (2017), \url{http://opendata.cern.ch/record/7719}

\bibitem{data-7721}
{CMS Collaboration} (2017), \url{http://opendata.cern.ch/record/7721}

\bibitem{data-7722}
{CMS Collaboration} (2017), \url{http://opendata.cern.ch/record/7722}

\bibitem{data-7723}
{CMS Collaboration} (2017), \url{http://opendata.cern.ch/record/7723}

\bibitem{data-9588}
{CMS Collaboration} (2017), \url{http://opendata.cern.ch/record/9588}

\bibitem{Agostinelli:2002hh}
S.~Agostinelli et~al. (GEANT4), Nucl. Instrum. Meth. \textbf{A506}, 250 (2003)

\end{thebibliography}

\end{document}